\documentclass[english,aps,pra,twocolumn,showpacs]{revtex4-1}
\usepackage{hyperref}
\hypersetup{
  pdftitle={Theoretical study of terahertz generation from atoms and aligned molecules driven by two-color laser fields},
  pdfauthor={Wenbo Chen, Yindong Huang, Chao Meng, Jinlei Liu, Zhaoyan Zhou, Dongwen Zhang, Jianmin Yuan, and Zengxiu Zhao*},
  unicode=true,
  colorlinks=true,
  allcolors=blue,
  pagebackref=true,
 }

\usepackage{amsmath,amssymb,amsxtra,amstext}
\usepackage{graphicx}
\usepackage{mhchem}
\newlength{\fntxvi} \newlength{\fntxvii}
\newcommand{\chemical}[1]
{{\fontencoding{OMS}\fontfamily{cmss}\selectfont
    \fntxvi\the\fontdimen16\font\
    \fntxvii\the\fontdimen17\font\
    \fontdimen16\font=3pt\fontdimen17\font=3pt
    \mathrm{#1}
    \fontencoding{OMS}\fontfamily{cmys}\selectfont
    \fontdimen16\font=\fntxvi \fontdimen17\font=\fntxvii}}
\usepackage{braket}
\makeatother

\begin{document}

\global\long\def\xkh#1{\left( #1 \right)}
\global\long\def\zkh#1{\left[ #1 \right]}
\global\long\def\dkh#1{\left\{  #1 \right\}  }
\global\long\def\rdkh#1{\left. #1 \right\}  }
\global\long\def\ldkh#1{\left\{  #1 \right.}
\global\long\def\no#1{#1{\!\!\!\backslash}}
\global\long\def\Doslash#1{#1\!\!\!\!\slash}
\global\long\def\doslash#1{#1\!\!\!\slash}
\global\long\def\dopartial#1#2{\frac{\partial#1}{\partial#2}}
\global\long\def\citenumber#1{{[\!\!\citenum{#1}]}}
\global\long\def\upcite#1{\textsuperscript{\cite{#1}}}
\global\long\def\Ket#1{\left| #1 \right\rangle }
\global\long\def\Bra#1{\left\langle #1 \right|}
\global\long\def\ket#1{|#1 \rangle}
\global\long\def\bra#1{\langle#1| }
\global\long\def\vbraket#1#2#3{\left\langle #1|#2|#3\right\rangle }
\global\long\def\vBraket#1#2#3{\left\langle #1\right|#2\left|#3\right\rangle }
\global\long\def\bracket#1#2{\left\langle #1 | #2 \right\rangle }
\global\long\def\vbracket#1#2#3{\left\langle #1 \right|#2 \left|#3 \right\rangle }
\global\long\def\vbracketn#1#2#3{\langle#1 |#2 |#3 \rangle}
\global\long\def\sech#1{\textmd{sech}#1}
\global\long\def\mo#1{\left| #1 \right|}
\global\long\def\mosq#1{\left| #1 \right|^{2}}
\global\long\def\dint#1{\!\! #1 \,}
\global\long\def\ddint#1{\!\! #1 \!}
\global\long\def\vect#1{\mathbf{#1}}

\preprint{}

\preprint{Draft \#1}

\title{Theoretical study of terahertz generation from atoms and aligned molecules driven by two-color laser fields}

\author{Wenbo \surname{Chen}}


\author{Yindong \surname{Huang}}

\author{Chao \surname{Meng}}

\author{Jinlei \surname{Liu}}

\author{Zhaoyan \surname{Zhou}}

\author{Dongwen \surname{Zhang}}

\author{Jianmin \surname{Yuan}}

\author{Zengxiu \surname{Zhao}}

\email{zhao.zengxiu@gmail.com}

\affiliation{Collage of Science, National University of Defense Technology, Changsha 410073, P. R. China.}

\date{\today}
\begin{abstract}
We study the generation of terahertz radiation from atoms and molecules driven by an ultrashort fundamental laser and its second harmonic field by solving time-dependent Schr\"odinger equation (TDSE). The comparisons between one-, two-, and three- dimensional TDSE numerical simulations  show that initial ionized wave-packet and its subsequent acceleration in the laser field and rescattering with long-range Coulomb potential play key roles. We also present the dependence of the optimum phase delay and yield of terahertz radiation on the laser intensity, wavelength, duration, and the ratio of two-color laser components. Terahertz wave generation from model hydrogen molecules are further investigated by comparing with high harmonic emission. It is found that the terahertz yield is following the alignment dependence of ionization rate, while the optimal  two-color phase delays varies by a small amount when the alignment angle changes from 0 to 90 degrees, which reflects alignment dependence of attosecond electron dynamics. Finally we show that terahertz emission might be used to clarify the origin of interference in high harmonic generation from aligned molecules by coincidently measuring the angle-resolved THz yields.
\end{abstract}


\keywords{Insert suggested keywords here --- APS authors don't need to do this.}

\maketitle

\begin{turnpage}

\end{turnpage}

\begin{ruledtabular}
\end{ruledtabular}

\begin{turnpage}

\end{turnpage}

\section{Introduction\label{sec:Introduction}}

Terahertz (THz) radiation, with wavelength between 0.03 to 3 millimeters, has been found useful for applications in information and communication technology, homeland security, global environmental monitoring and ultrafast computing, to name a few \cite{Tonouch07,Hoffmann07}. Scientifically, it provides resonant access into and hence enables the probing of various of modes, such as the motion of free electrons, the rotation of molecules, the vibration of crystal lattices and the precessing of spins \cite{Kampfrath13Review}. With power continually being increased and duration being shorten into single or half cycle, pulsed intense THz sources are now capable of driving and steering non-resonant ultrafast processes in matter in unique ways. New phenomena are being observed, e.g., THz harmonic generation \cite{Zaks12}, strong-field control of spin excitation \cite{Kampfrath2011} and phase-transition \cite{Liu2012j}, etc.

The efficient generation of THz wave, however, is found still challenging. The popular approaches include optical THz generation, solid-state electronics and quantum cascade laser, which are limited either by low frequency conversion efficiency, low temperature requirement, or the lack of eigen oscillator \cite{Saeedkia08}.
Among these techniques, laser air photonics is capable of generating THz field strength greater than 1 MV/cm with bandwidth of over 100 THz through plasma formation by laser ionized gaseous medium \cite{Clough12}. But the mechanism of THz generation from the laser induced plasma turns out be complicated \cite{Thomson07} involved of wave propagation, plasma formation, oscillation and collision. For example, the THz can be emitted by the radical acceleration of the ionized electrons due to the ponderomotive force generated by the radial intensity gradient of the optical beam \cite{Hamster93}, or by linear mode conversion from the laser-induced plasma wakefield in inhomogeneous plasma \cite{Sheng04, Sheng05}, or by Cerenkov-type mechanism from light propagating within the filament \cite{DAmico07L}.

In this work, we restrict our study to single atom response by focusing on THz generation from gases of much lower densities in low pressure and minimizing the collective motion of plasma and its modification on the light propagation. Even in this much simpler case, the mechanism of THz emission is still under debated, partly because of the involved complex dynamics during laser interaction with gases. Nevertheless, down to the fundamental origin of THz emission, it could only either be the induced current or the induced polarization. Therefore two popular mechanism are proposed, a microscopic photocurrent (PC) model with tunneling photoionization \cite{Kim07, Kim08NP} and a third-order nonlinear model with four-wave mixing (FWM) \cite{Cook00,Xie06,Dai06}.

Both models qualitatively agree with the experimental findings that the broken symmetry of the laser-gas interaction could enhance the field strength of the emitted THz waves by several order-of-magnitudes, when either applying a DC bias \cite{Loffler00, Loffler02}, an AC-bias by using two-color pulses \cite{Cook00} or using carrier-envelop phase stabilized few-cycle pulse \cite{Kress06}.
As a representative case, THz emission in two-color laser fields has been investigated more intensively \cite{Cook00,Kress04,Kress06,Xie06,Gildenburg07,Kim07,Kim08NP,Wen09,Dai09,Babushkin10,Babushkin11,Berge13}. So far, more evidences support the model based on photocurrent as it is found ionization is indispensable \cite{Cook00}, although the physics behind photocurrent formation needs to be explored carefully \cite{ZhangDW12}. However, in view of the success of four-wave mixing model in applying to the THz detection through
air-breakdown coherent detection scheme \cite{Dai06, Lv13}, the two models must be closely connected.

The full quantum description in our previous work and others \cite{ZhouZY09,Silaev09,Karpowicz09,ZhangDW12} provides more insight into the mechanism of THz emission. By connecting THz emission with high harmonic generation (HHG) in two-color laser pulses, the dynamics responsible for THz emission is attributed to the continuum-continuum transition during laser-assisted soft collision of the ionized electron with the atomic core, while the hard collision with the atomic core leads to HHG upon recombination \cite{ZhangDW12,Lv13}.

Because the fundamental pulse is strong enough to induce ionization, its interaction with the gas can not be treated perturbatively. The interaction with the second harmonic pulse is however weak which seems to be possibly treated perturbatively on the top of the non-perturbative behavior from the fundamental pulse. In this sense, the photocurrent model can be understood by another type of four-wave mixing model, but differed from the usual case that the involved lower and up states could be really populated by multiphoton ionization or tunneling ionization. Therefore the two models can be unified as hinted in \cite{Thomson07, ZhouZY09,ZhangDW12,Lv13}.

Although the second harmonic pulse is very weak, it serves as a gate controlling the phase of the electron wavepacket and hence breaks the left to right symmetry leading to the formation of net current and the emission of THz waves. At the same time, the out of phase high harmonic emission between adjacent half cycle in single color pulse is changed as well leading to the emission of even-order high harmonics \cite{Dudovich06N, Mauritsson06, Oishi06,Doumy09,Dahlstrom11}. Therefore, by controlling the phase delay between the fundamental and the second harmonic pulse, the THz and high harmonic yield can be controlled simultaneously. This enables the calibration of the optimal phase delay (OPD) maximum THz yields and it is found that the OPDs related to the atomic potential \cite{ZhangDW12,Lv13}. It is thus expected that the sensitivity of THz yields to the Coulomb potential might be used to map atomic fields from within and to help the full characterization of the rescattering wave packet \cite{ZhangDW12}.

In the present investigation, we explore further the effects of atomic potential and other laser parameters on the generation of THz for both a hydrogen atom and a model H$_2^+$ molecule interacting with two-color pulses. The paper is organized as follows. In section II, we present the theory and model for THz generation from ionizing atoms by two-color laser pulses. In section III, we show the calculated results and discuss the physics behind. Finally we draw conclusions in section IV.

\section{Theory and methods\label{sec:Theory}}

\subsection{Photocurrent model}

Under strong laser fields, the electron current is presumably contributed by photoelectrons released after ionization. When the laser photon energy is much less than the ionization potential $I_p$, and $I_p<2U_p$ where the ponderomotive energy $U_p$ is the averaged kinetic energy of an electron moving in the laser field, the mechanism of ionization can be considered as tunneling of electron through the potential barrier formed by the atomic potential and the instantaneous electric dipole potential. The instant ionization rate in the electric field of $E(t)$ can be obtained using Ammosov-Delone-Krainov (ADK) model \cite{Ammosov86,Priori2000}
\begin{equation}
w\!\left(t\right)=I_{p}\left|C_{n^{*}}\right|^{2}\times\left\{ \frac{2\kappa^{3}}{\left|E\!\left(t\right)\right|}\right\} ^{2n^{*}-1}\exp{\!\left\{ -\frac{2\kappa^{3}}{3\left|E\!\left(t\right)\right|}\right\} }\label{eq:ADKIonRate}
\end{equation}
where $Z_{c}$ and $I_{p}$ are the net resulting charge and ionization potential of the atom, $I_{ph}=0.5\mbox{au}$ is the ionization potential of the hydrogen atom, respectively. And $\kappa=\sqrt{2I_{p}}$, $n^{*}=Z_{c}\sqrt{I_{ph}/I_{p}}$, $C_{n^{*}}=\frac{2^{2n^{*}}}{n^{*}\Gamma\left(n^{*}+1\right)\Gamma\left(n^{*}\right)}$.

If the laser field strength is much higher such that the binding energy of the electron is higher than the height of the potential barrier, ionization can be proceed through over-the-barrier ionization (OTBI). In OTBI range, we can use the following empirical formula suggested by \cite{Tong05},
\begin{equation}
w_{OTBI}\!\left(t\right)=w_{TI}\!\left(t\right)e^{-\alpha\left(Z_{c}^{2}/I_{_{p}}\right)\left(E\!\left(t\right)/\kappa^{3}\right)},\label{eq:OTBIIonRate}
\end{equation}

Once the instantaneous ionization rate is found, the free electron density as a function of time can be obtained by
\begin{equation}
n_{e}\!\left(t\right)=n_{0}\left\{ 1-\exp\!\left[-\int_{-\infty}^{t}\!\! w\!\left(t'\right)dt'\right]\right\} ,\label{eq:ADKEelectronDensity}
\end{equation}
where $n_{0}$ is the neutral atom density. Assuming the subsequent motion of the electron after ionization is dominated by the laser field, the photon-current and its acceleration contributed by the ionization event $t_{i}$ are
\[
\begin{aligned}
j\!\left(t_{i},t\right)= & -n_{0}\!\left(t_{i}\right)w\!\left(t_{i}\right)\left[p\!\left(t_{i}\right)+\int_{t_{i}}^{t}\!\! E\left(t'\right)dt'\right],\\
a\!\left(t_{i},t\right)= & -n_{0}\!\left(t_{i}\right)w\!\left(t_{i}\right)E\!\left(t\right).
\end{aligned}
\]
Note that we take the initial longitudinal momentum $p\!\left(t_{i}\right)=0$ at the instant ionization. The total photon-current $J\!\left(t\right)$ and dipole acceleration $A\!\left(t\right)$ are thus formulated as
\[
\begin{aligned}J\!\left(t\right)= & \int_{-\infty}^{t}\!\! j\!\left(t_{i},t\right)dt_{i},\\
A\!\left(t\right)= & \int_{-\infty}^{t}\!\! a\!\left(t_{i},t\right)dt_{i}.
\end{aligned}
\]

From the current, the net residual current density (RCD) after the pulse is given by $J_{RCD}=J\!\left(t\rightarrow\infty\right)$.
The frequency-domain spectrum $F\!\left(\Omega\right)$ are calculated by the Fourier transformation of acceleration $A\!\left(t\right)$ with intensity given by $G(\Omega)=\mosq{F\xkh{\Omega}}$. The total yields of THz radiation is given by the integral of the spectra intensity from 0 up to 30 THz.

\subsection{Time-dependent Schr\"odinger equation}

The time-dependent Schr\"odinger equation (TDSE) for an atom interaction with the laser field in dipole approximation is given by
\begin{equation}
i\frac{\partial}{\partial t}\Psi\xkh{\vect r,t}=\left[\frac{\hat{\vect p}^{2}}{2}+\hat{V}\xkh r+\vect r\cdot\vect E(t)\right]\Psi\xkh{\vect r,t}
\end{equation}
where the linearly polarized (along $\hat{z}$ direction) two-color laser field is given by
\begin{equation}
\vect E(t)=\left[E_{\omega}\cos\left(\omega t\right)+E_{2\omega}\cos\left(2\omega t+\phi\right)\right]\hat{z}f(t). \label{eq:elasert}
\end{equation}
Here, $E_{\omega}$, $E_{2\omega}$, $f(t)$, $\omega$, $\phi$ denote fundamental and its second-harmonic laser field strength, temporal envelop, fundamental frequency, two-color phase delay, respectively.
By numerically solving the TDSE in three-dimension, the induced dipole acceleration along $\hat{z}$ direction is given by the quantum average:
\begin{equation}
a(t)=\langle\frac{\partial V}{\partial z}-E(t)\rangle,
\end{equation}
from which the THz radiation can be obtained by Fourier transformation.
The corresponding electron current density \cite{Silaev10} is given by $J(t) =\int_{-\infty}^{t}{a}\left(t\right)dt$ and can be recast into
\begin{align*}
J(t) & =J'(t)+J_{bb}(t)
\end{align*}
where the fast oscillation part $J_{bb}(t)$ results from the dipole transitions between the populated Rydberg states, and $J'(t)$ is the rest part of the current, which generates the broadband terahertz radiation. The residual current density can be found numerically by time-averaging the current after the laser turns off to cancel the $J_{bb}(t)$ component, $J_{RCD}=\frac{1}{\tau}\int_{t_{f}}^{t_{f}+\tau}\dint{dt}J(t)$ which can be directly compared to that from the PC model. The THz spectra and yields can be computed from the dipole acceleration at the same manner given in the previous subsection.

\begin{figure}[htdp]
\includegraphics[clip,width=0.8\columnwidth]{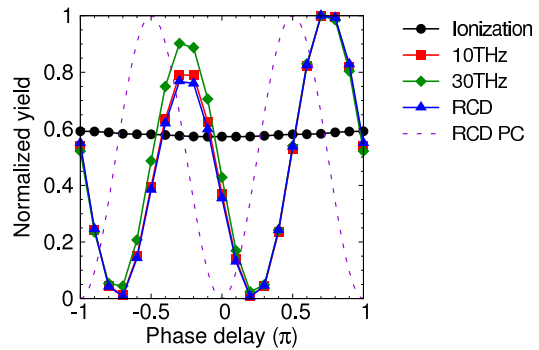}
\caption{\label{fig:thzwi_3d34_2400tt_a}(color online) Modulation of THz yields, residual current density and ionization probability as functions of the phase-delay between the two color pulses. The TDSE calculation is carried out for hydrogen atoms. The fundamental laser pulse has the intensity of $2\times10^{14}\mbox{W/cm}^{2}$, the wavelength of 800nm, the FWHM of 25fs, and the second harmonic intensity is 1\% of fundamental pulse.}
\end{figure}

As an example, we show in Fig. \ref{fig:thzwi_3d34_2400tt_a} the typical THz yield modulations from an Hydrogen atom in the two-color laser pulse, obtained from the TDSE simulation. As the two-color phase delay varies, the modulations of terahertz yield below 10THz, 30THz and RCD takes the maximum at nearly the same optimum phase delay (OPD) $\phi_{m}$ as shown in Fig. \ref{fig:thzwi_3d34_2400tt_a}. It suggests that the RCD can be used as a good parameter characterizing the phase-delay dependence of THz yields. On the other hand, the ionization probability is less modulated by the applied weak second harmonic pulse which indicates that the modulation of THz yield is mainly contributed by the controlling of the electron dynamics after ionization in accordance to the analysis presented in \cite{ZhangDW12, Lv13}. For comparison, we show the modulation of RCD from PC model, which deviates from the TDSE calculation because of the neglecting of the Coulomb potential in the dynamics of free electrons (???).

\section{Results and discussions\label{sec:Results}}

\subsection{Laser intensity, wavelength and atomic orbitals}

\begin{figure}[htdp]
\includegraphics[clip,height=0.47\columnwidth]{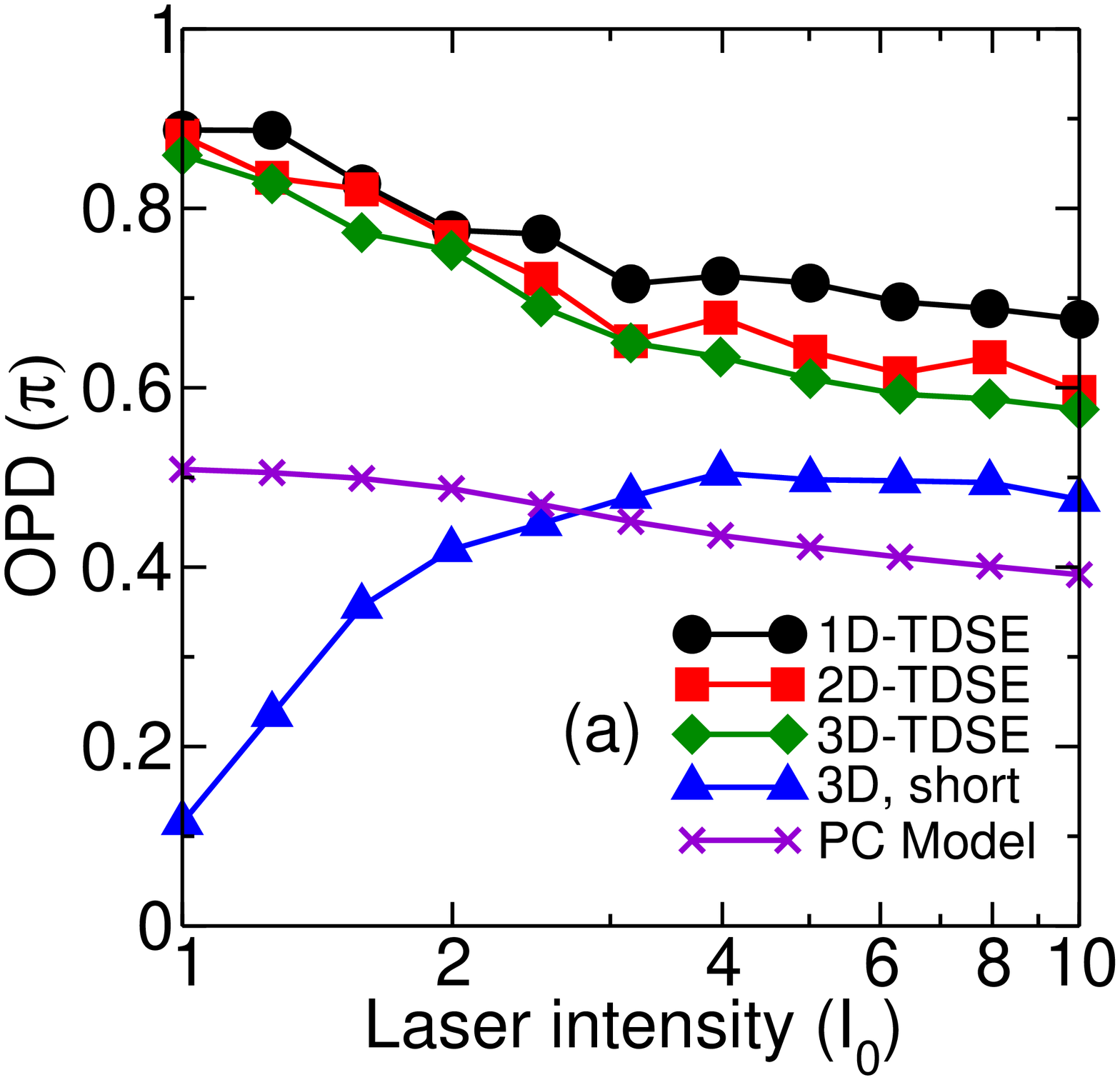}
\includegraphics[clip,height=0.47\columnwidth]{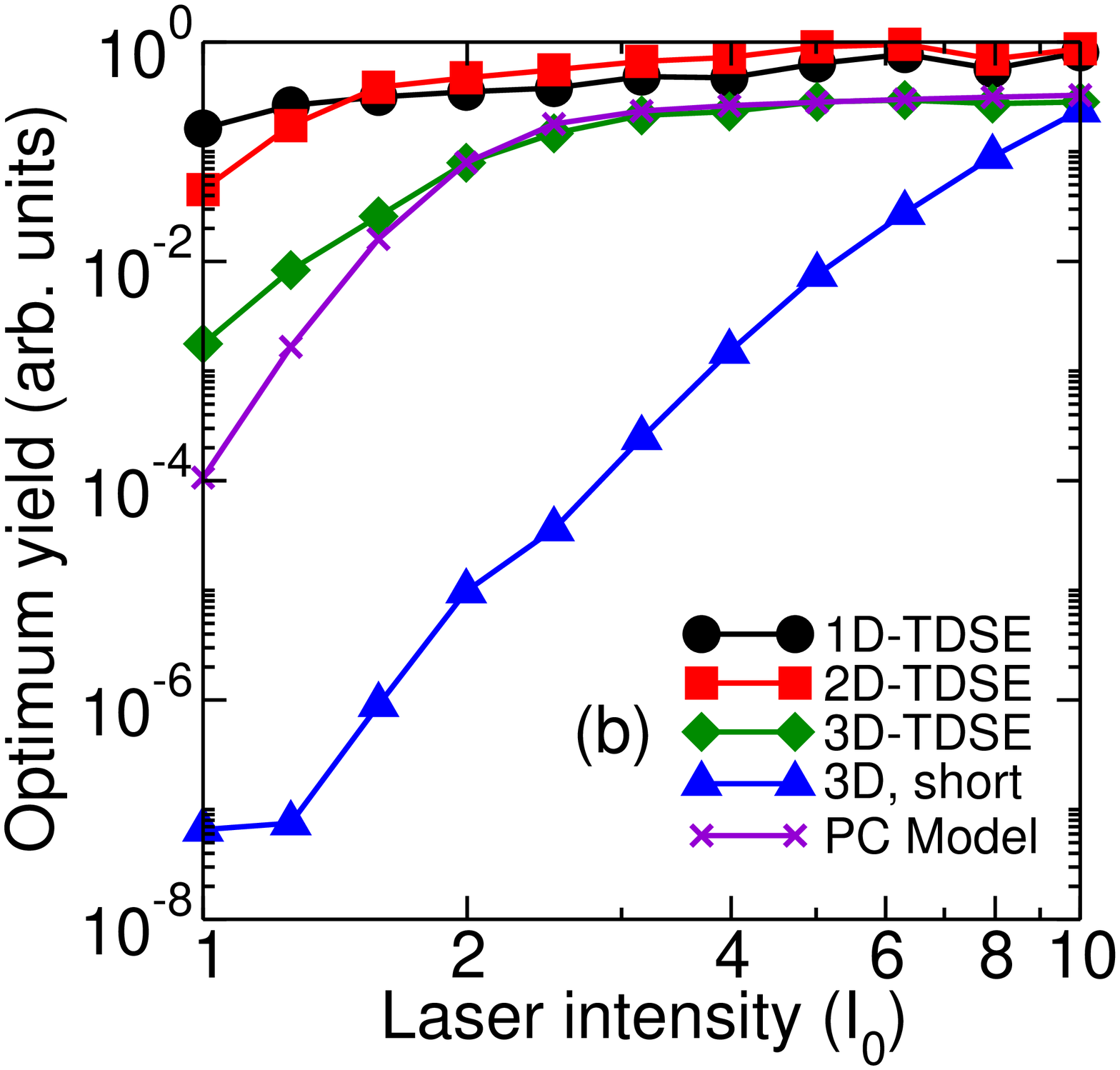}
\caption{\label{fig:phad-1d2d3d24-1100-1}(color online) The optimal yields (a) and the corresponding optimal phase-delay (b) of THz generation at various of laser intensities calculated by 1D- (black circle), by 2D- (red square), by 3D- (green diamond) TDSE with Coulomb potential, 3D-TDSE short range potential. The laser intensity is in unit of $I_0$ with  $I_{0}=1\times10^{14}\mbox{W/cm}^{2}$, and the ionization potential is $I_p=0.5$ atomic units for all cases.}
\end{figure}

We first examine in details how the THz yields from hydrogen atoms depend on various of parameters of the two-color laser pulses, especially we pay attention to the optimal phase-delay that maximizes the THz yields. In Fig. \ref{fig:phad-1d2d3d24-1100-1}, the laser-intensity dependence of the optimal phases and the corresponding THz yields are compared between different models. The ionization potential is adjusted taking the same value of $I_p=0.5$au. When the atomic potential is the long-range Coulomb potential, for either 1D, 2D or 3D calculations, the optimal phase follows the same trend that changes from 0.9$\pi$ to 0.6$\pi$ as the laser intensity varies from $I_0$ to $10I_0$ where $I_0=10^{14}$W/cm$^2$. In contrast, the optimal phase predicted by the PC model varies little around 0.5$\pi$ (the variation might be due to the depletion of the ground state). It confirms our previous investigation that PC model for THz generation fails due to the ignorance of the Coulomb potential \cite{ZhangDW12}.  When a short range potential rather than the Coulomb potential is used, such as for the negative ions, the optimal phase of THz generation shows complete different behavior, increasing from 0.1$\pi$ to $0.5\pi$ as the laser intensity increases as shown in Fig.\ref{fig:phad-1d2d3d24-1100-1}(a). It clearly demonstrates the importance of the long range potential.

The variation of the OPD with laser intensity indicates that different ionization mechanism are involved.
Let us now focus on the optimal phases from the 3D calculations with Coulomb potentials that are closely related to the experimental observation. As the laser intensity increases, the ionization mechanisms changes from multi-photon ionization (MPI) to tunneling ionization (TI) and then over-the-barrier ionization (OTBI).

In the regime of MPI when $\gamma>1$, terahertz generation is dominated by four-wave-mixing \cite{ZhouZY09} with prediction of the terahertz field strength given by $E_{THz}\left(t\right)\propto\chi^{(3)}E_{2\omega}\left(t\right)E_{\omega}^{*}\left(t\right)E_{\omega}^{*}\left(t\right)\cos\phi$ and the OPD is close to 0 or $\pi$, in agreement with the numerical results at the laser intensity of $I_0=10^{14}$W/cm$^2$.
When TI dominate ($\gamma<1$), the soft-recollision between ionized electron with atomic core's Coulomb potential play the key role \cite{ZhangDW12, Lv13}. Rescattering currents dominate and escaped currents vanish, and the OPD shifts from $\pi$ to $0.6\pi$.
As further increasing of the laser intensity, OTBI takes over for laser intensity over $I=\frac{\kappa^{4}}{16I_{p}}$. The non-vanishing initial longitudinal momentum cause the enhancing of the escaped part of the electron wave packet, and hence the OPD is close to $0.5\pi$. In this case, the atomic potential might be neglected and hence the photon-current model holds predicting the terahertz field as $E_{THz}\left(t\right)\propto dJ\left(t\right)/dt\propto f\left[E_{\omega}\left(t\right)\right]E_{2\omega}\left(t\right)\sin\phi$.

We now turn to the dependence of optimal terahertz yields (OTY) on the laser intensity. Surprisingly, the 3D calculation and the photocurrent model give quite similar dependence. It indicates that the ionization rate is crucial in determining the yields, since the ADK ionization rate is derived for 3D Hydrogen atom. In contrast, the 1D and 2D calculations over-estimate the ionization yield, because the wave packet is less spreading comparing the 3D case. It is also worth noted that the yields from the short range potential are much lower for two reasons. First the tunneling ionization rate at the same ionization energy is indeed smaller for short range potential comparing to the long range potential. Secondly, the laser-assisted soft collision is suppressed in the case of short range potential.

\begin{figure}[htdp]
\includegraphics[clip,height=0.47\columnwidth]{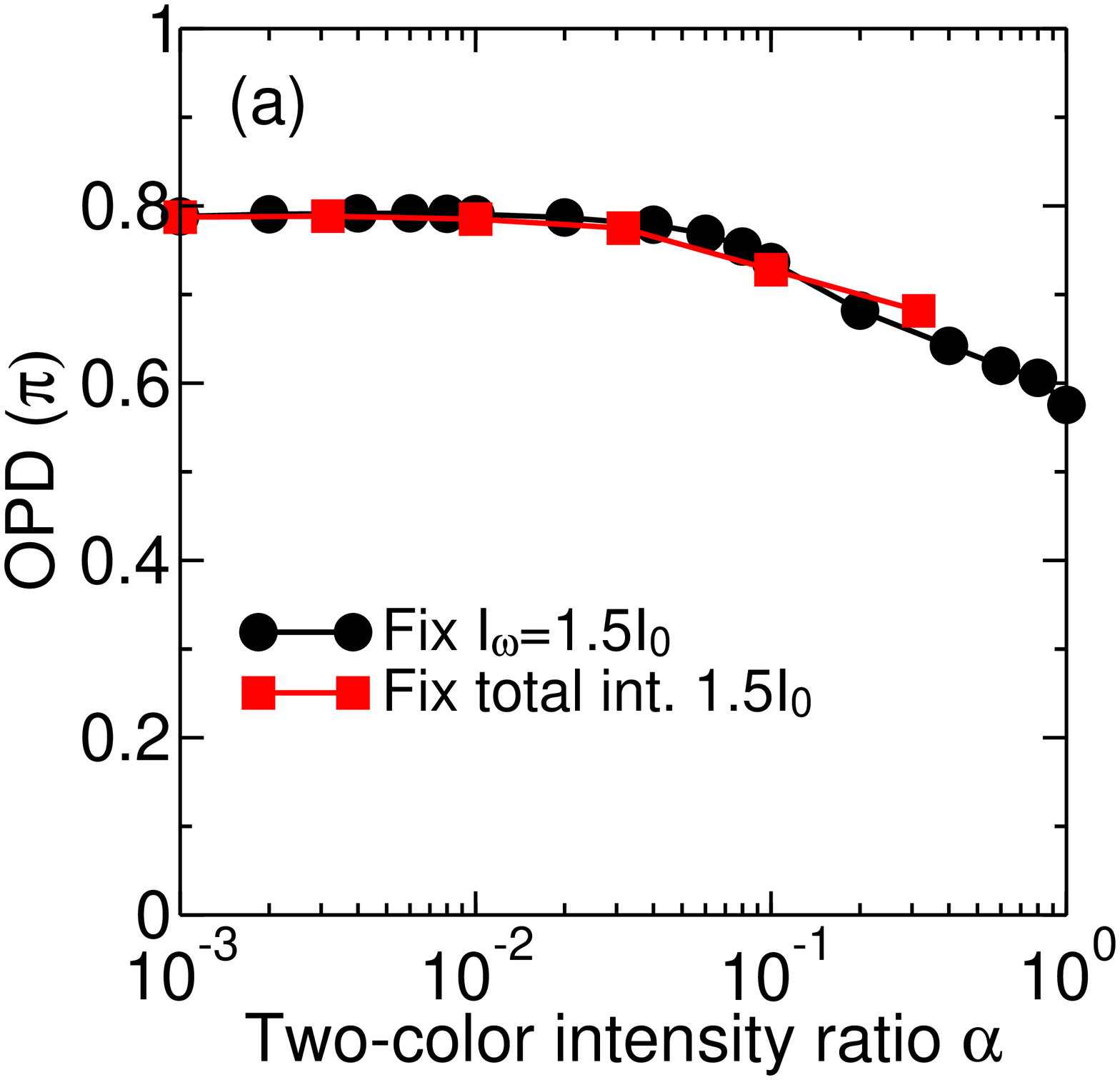}
\includegraphics[clip,height=0.47\columnwidth]{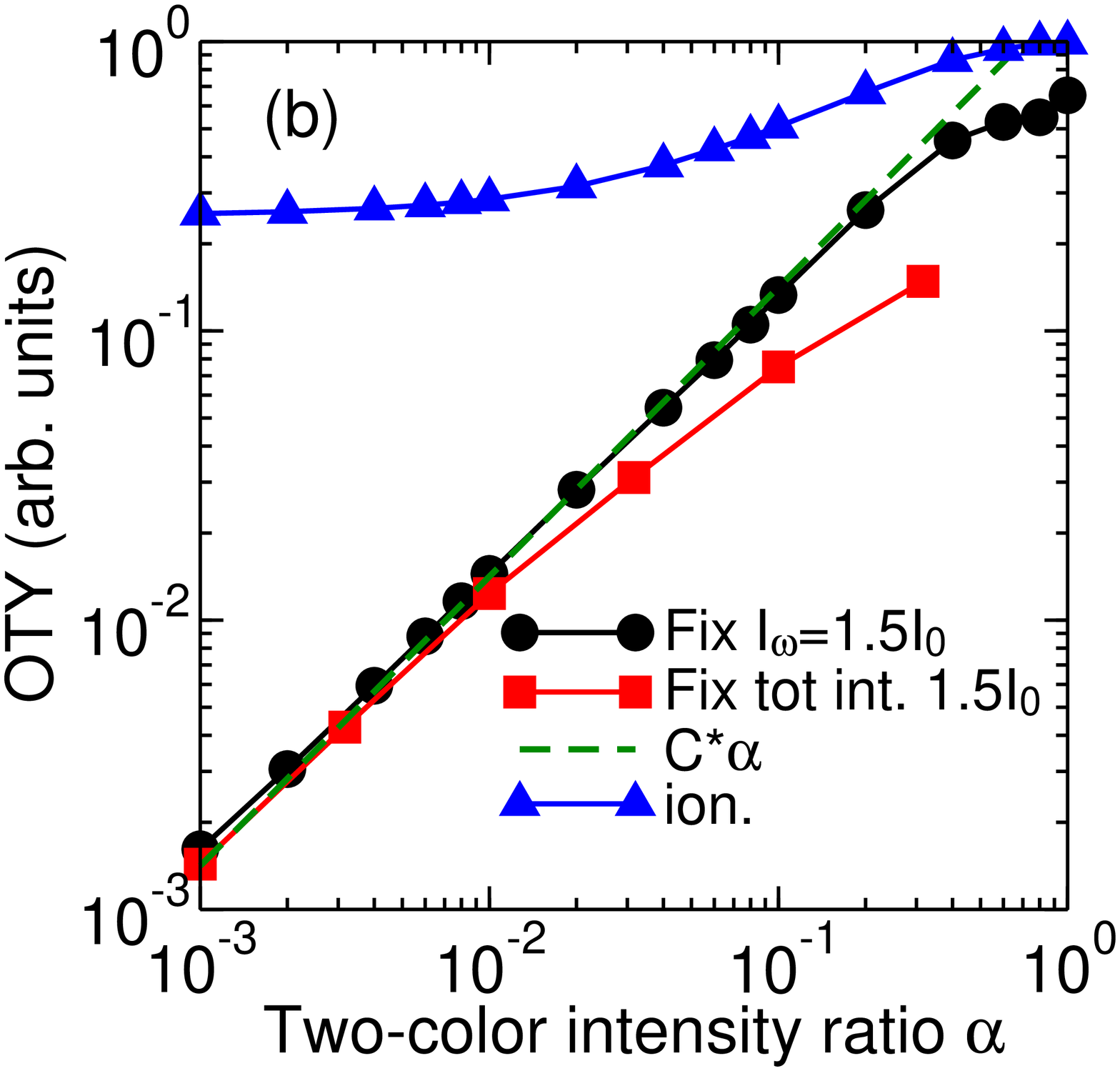}
\caption{\label{fig:2w1w-3d3541jrcd2}(color online) The scaling of OPD (a) and OTY (b) of THz generation with the intensity ratio of the second harmonic pulse to the fundamental laser pulse.}
\end{figure}

Next, we examine how the optimal phase delay and yields depend on the intensity ratio, $\alpha=I_{2\omega}/I_{\omega}=E_{2\omega}^2/E_{\omega}^2$, between the second harmonic and the fundamental pulse.  As shown in Fig.\ref{fig:2w1w-3d3541jrcd2}(a), the optimal phase delay remains nearly constant for $\alpha<0.1$ for given fundamental laser intensity ($I_{\omega}=1.5\times10^{14}\mbox{W/cm}^{2}$) of the fundamental pulse. But as the intensity ratio increases further, the OPD quickly drops to $0.5\pi$, because the second harmonic pulse begins to participate in the ionization process which eventually take over the ionization from the fundamental pulse. It is shown in Fig. \ref{fig:2w1w-3d3541jrcd2}(b) that the OTY's are scaled as power of the intensity ratio, which is consistent with the experiment \cite{Vvedenskii14}. The dependence on the intensity ratio of both the OPD and OTY are consistent with laser-assisted soft-collision model presented in \cite{Lv13}. For convenience of direct comparison to experiments, we present the intensity-ratio dependence by fixing the total power of the fundamental and the second harmonic pulse and the similar conclusion can be drawn as well. The ionization probability is also shown in Fig. \ref{fig:2w1w-3d3541jrcd2}(b) which is less dependent on the intensity ratio, suggesting the dynamics following ionization is crucial in terahertz wave generation.

\begin{figure}[htdp]
\includegraphics[clip,height=0.45\columnwidth]{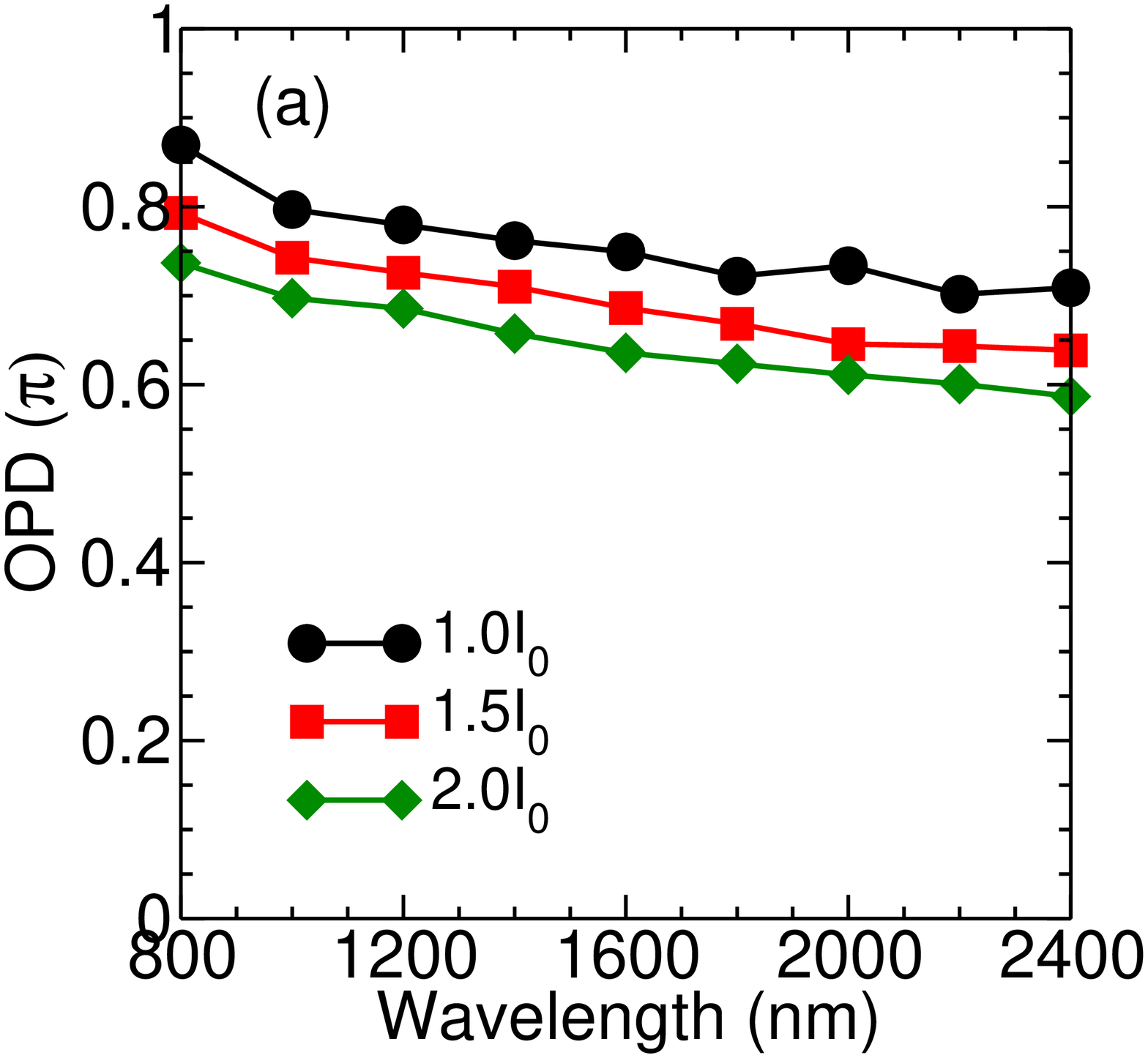}
\includegraphics[clip,height=0.45\columnwidth]{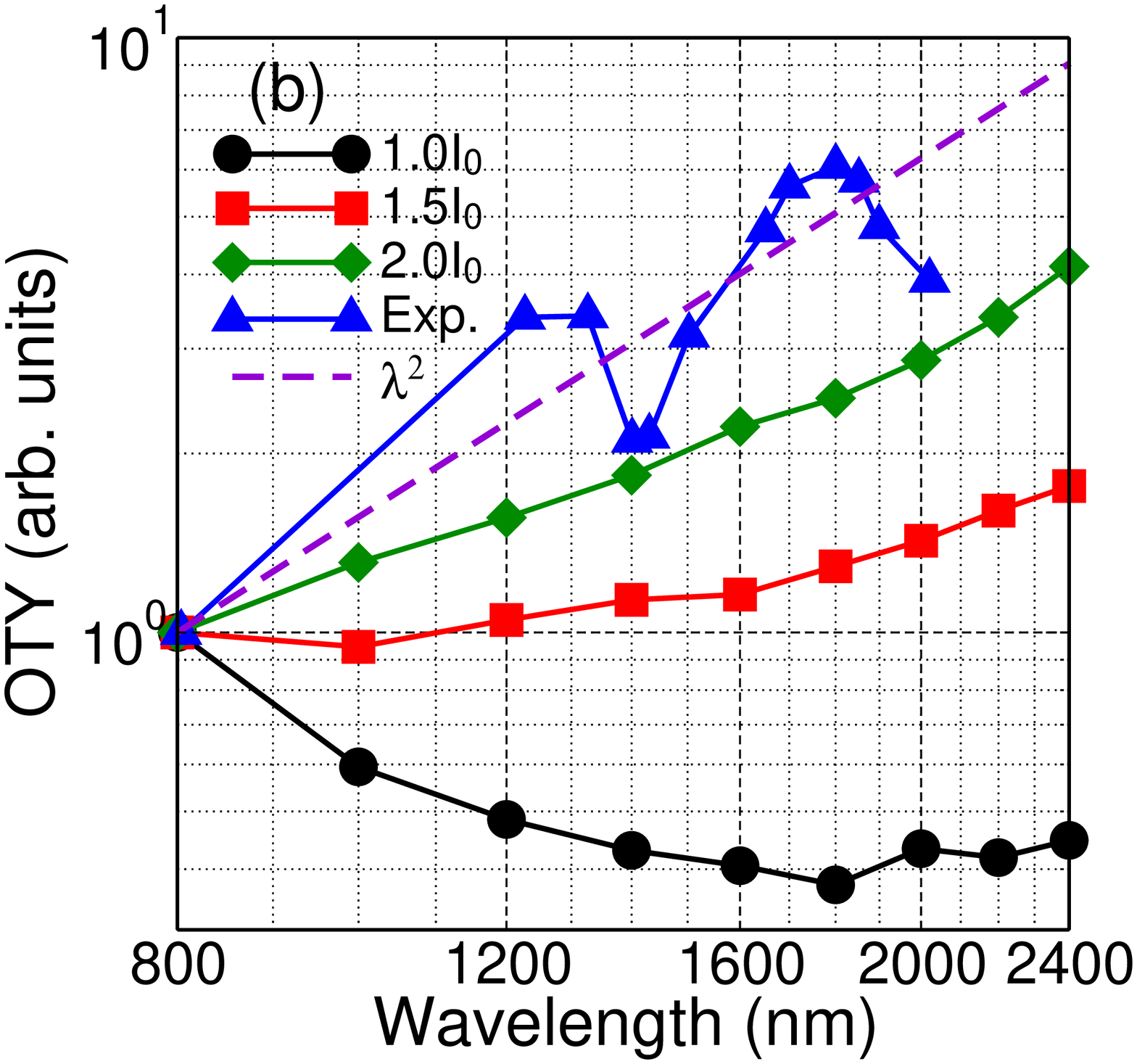}
\caption{\label{fig:wavlen-3d3271jrcd2-1}(color online) Wavelength scaling of (a) the optimum phase delay and (b) THz yields. The experimental data is taken from  \cite{Clerici2013}. Here, the two-color intensity ratio $\alpha=1\%$ and laser pulse has duration of FWHM 50fs.}
\end{figure}

To further clarify the mechanism of THz wave generation by two-color laser pulses, we examine the wavelength scaling of THz generation.
As the wavelength increases, the OPDs obtained from TDSE shift from $0.8\pi$ to $0.5\pi$ shown in Fig. \ref{fig:wavlen-3d3271jrcd2-1}(a), while the PC model predicts the OPD at $0.5\pi$ independent of wavelength. The variation of OPDs with wavelength might be rationalized as follows. Because the wave packet is more diffused during each cycle for the longer wavelength, the soft-recollision with atomic nuclear plays less rule resulted in the OPD at $0.5\pi$ as predicted by PC model, which neglects the Coulomb potential completely.

On the other hand, according to the PC model, the OTY's are scaled quadratically with wavelength, while the TDSE simulation shows deviation from this dependence. As shown in Fig. \ref{fig:wavlen-3d3271jrcd2-1}(b), the OTYs from the TDSE calculation are scaled as $\lambda^{1.45}$, at the laser intensity of $I_\omega=2I_0$. While for $I_\omega=1.5I_0$ or $I_0$ the deviation is even bigger. One of the reason for the deviation is the diffusion of the electron wave packet. However, the experimental data from Clerici et al \cite{Clerici2013} is more close to the predication by PC model. This discrepancy needs further investigation in the future.

\subsection{Terahertz wave generation from aligned molecules}

The previous investigation of THz generation from atoms in two-color laser fields demonstrates the importance of Coulomb potential on the laser-driven electron dynamics. In addition it confirms the sensitivity of THz yields on the detuning of ionization and electron dynamics. For molecules, it is therefore expected that THz generation might exhibit dependence on both molecular alignment and molecular potential.

To simulate molecular terahertz generation, we numerically solve the two-dimensional TDSE for a model $\chemical{H_{2}^{+}}$ with the soft-core potential between the electron and the two nuclei,
\[
V(\vect r)=-\frac{1}{\sqrt{\left(\vect r-\vect R/2\right)^{2}+a^{2}}}-\frac{1}{\sqrt{\left(\vect r+\vect R/2\right)^{2}+a^{2}}},
\]
where $\vect r=(z,x)$ and $\vect R=(R\cos\theta,R\sin\theta)$ are the electronic and molecular coordinates, $R$ is the internuclear distance. Two-color pulse is used. The fundamental and the second harmonic pluses are chose parallel in polarization along $\hat{z}$ axis. $\theta$ is the alignment angle between the molecular axis and the laser polarization $\hat{z}$, $a$ is the soft-core parameter. Here, the motion of the nuclei are frozen and we set $R=2\mbox{au}$, $a=1\mbox{au}$, and the ionization energy $I_{p}$ is equal to $0.934\mbox{au}$.

\begin{figure}[htdp]
\includegraphics[clip,height=0.45\columnwidth]{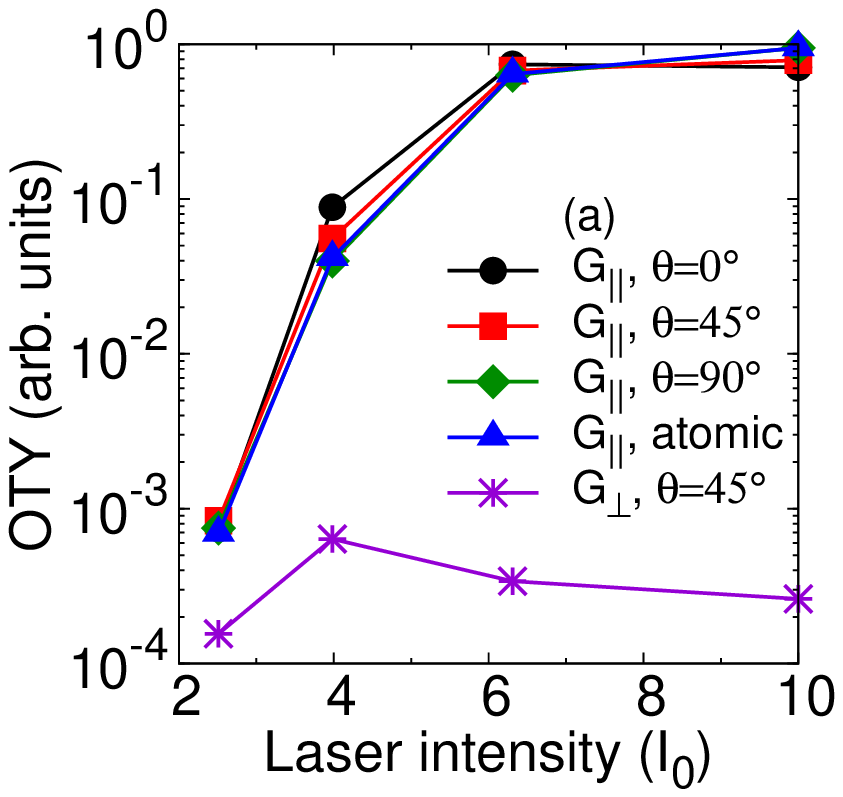}
\includegraphics[clip,height=0.45\columnwidth]{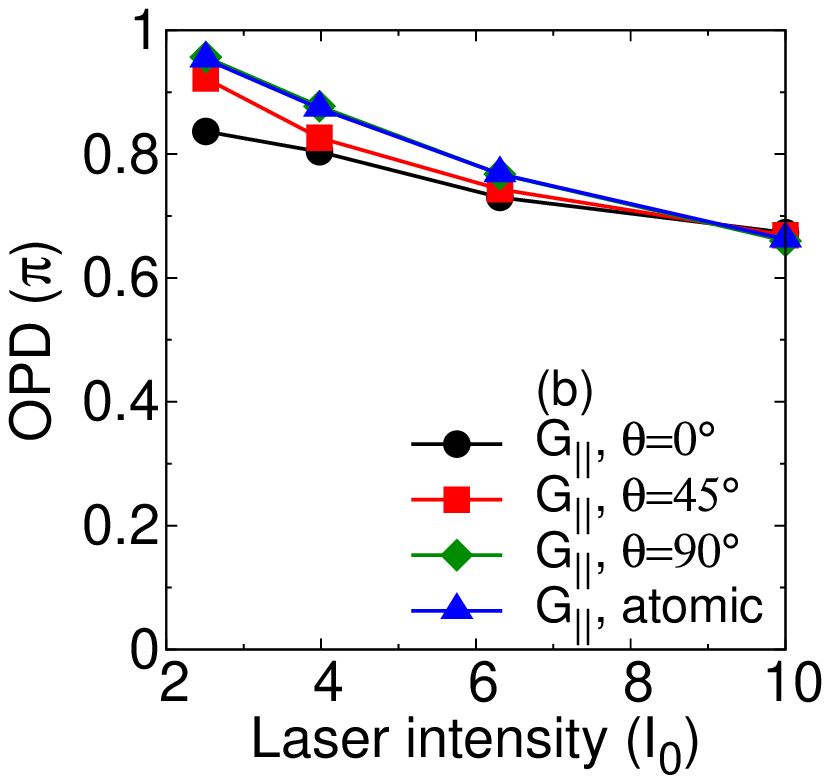}
\caption{\label{fig:molthz-opd-oty}(color online) Comparison of (a) the OTY and (b) the OPD of THz generation from molecules at different alignment and from atoms.}
\end{figure}

The emitted THz radiation are calculated from the low frequency part of the Fourier transformation of the induced dipole moments both parallel and perpendicular to the laser polarization. The total terahertz yield is the sum of yield of parallel $G_{\parallel}$ and perpendicular $G_{\perp}$ direction. As shown in figure \ref{fig:molthz-opd-oty}(a), when the laser intensity is in the tunneling regime ($\gamma>1$, and $I>2.1\times10^{14}\mbox{W/cm}^{2}$), the OTY of terahertz field parallel to the laser polarization is two orders larger than that from the perpendicular direction for molecules aligned at $45^\circ$ with respect to the laser polarization. At $0^\circ$ and $90^\circ$ alignment, the perpendicular components vanish. Therefore the total THz yields are dominated by the emission of THz wave with field component parallel to the laser polarization and in the follows we make no distinguishing between the total yield and the yield parallel to the laser polarization. Figure \ref{fig:molthz-opd-oty}(b) shows the OPDs of THz for molecules at different alignment. The OPD varies from $\pi$ in the MPI range to $0.6\pi$ in the TI or OTBI range as the laser intensity increases, similar to the case of atoms.

\begin{figure}[htdp]
\includegraphics[clip,height=0.31\columnwidth]{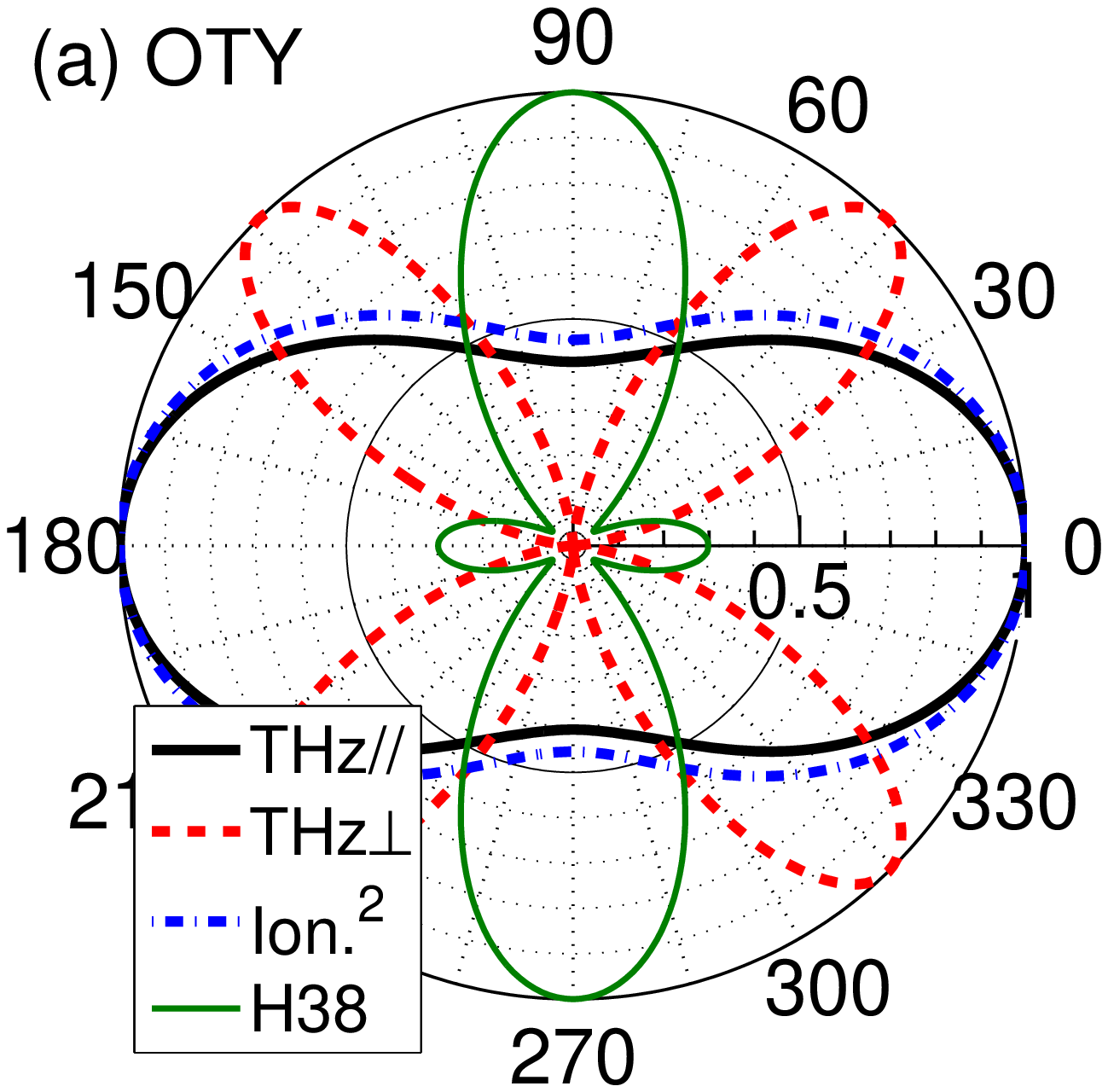}
\includegraphics[clip,height=0.31\columnwidth]{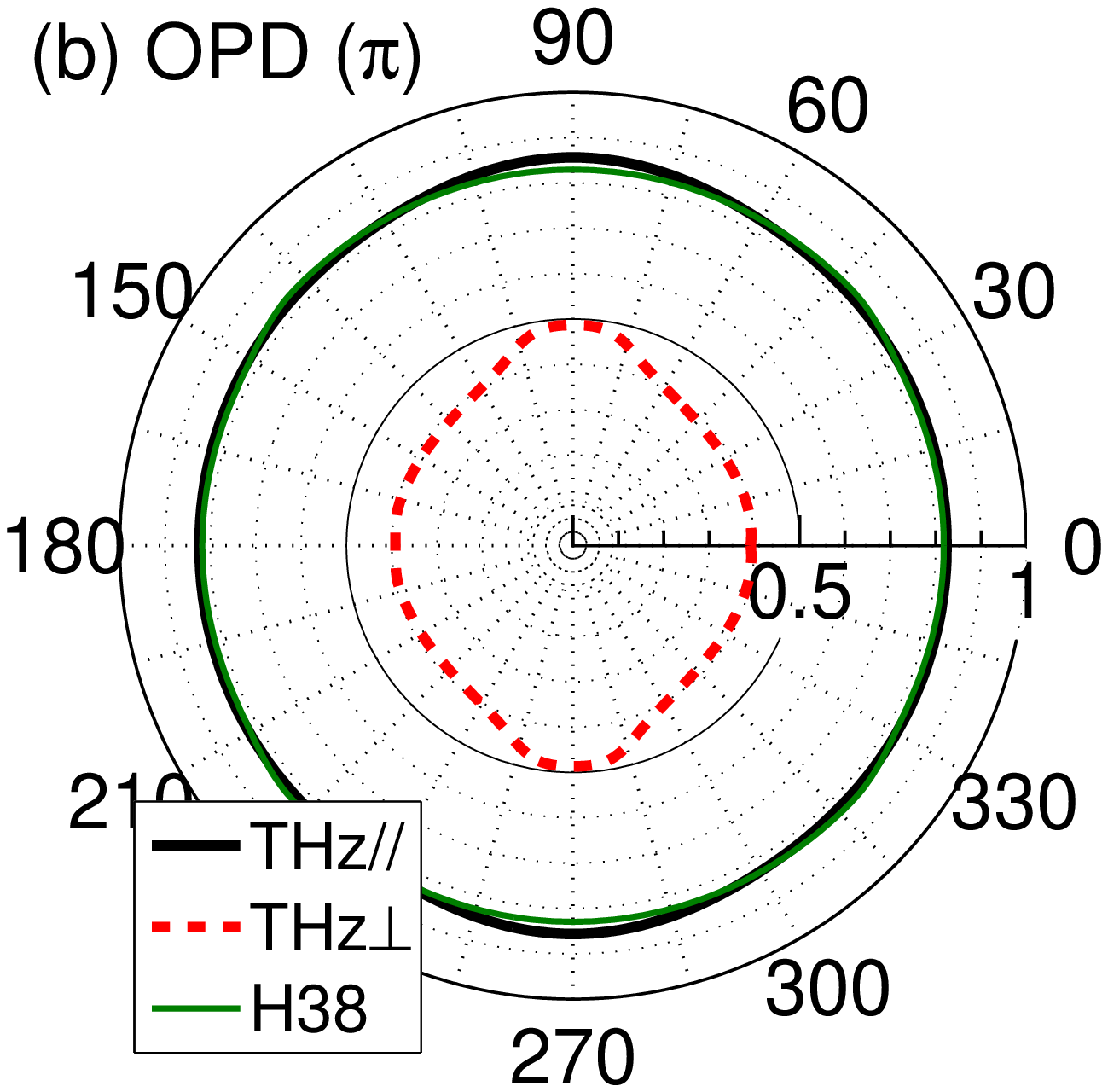}
\includegraphics[clip,height=0.31\columnwidth]{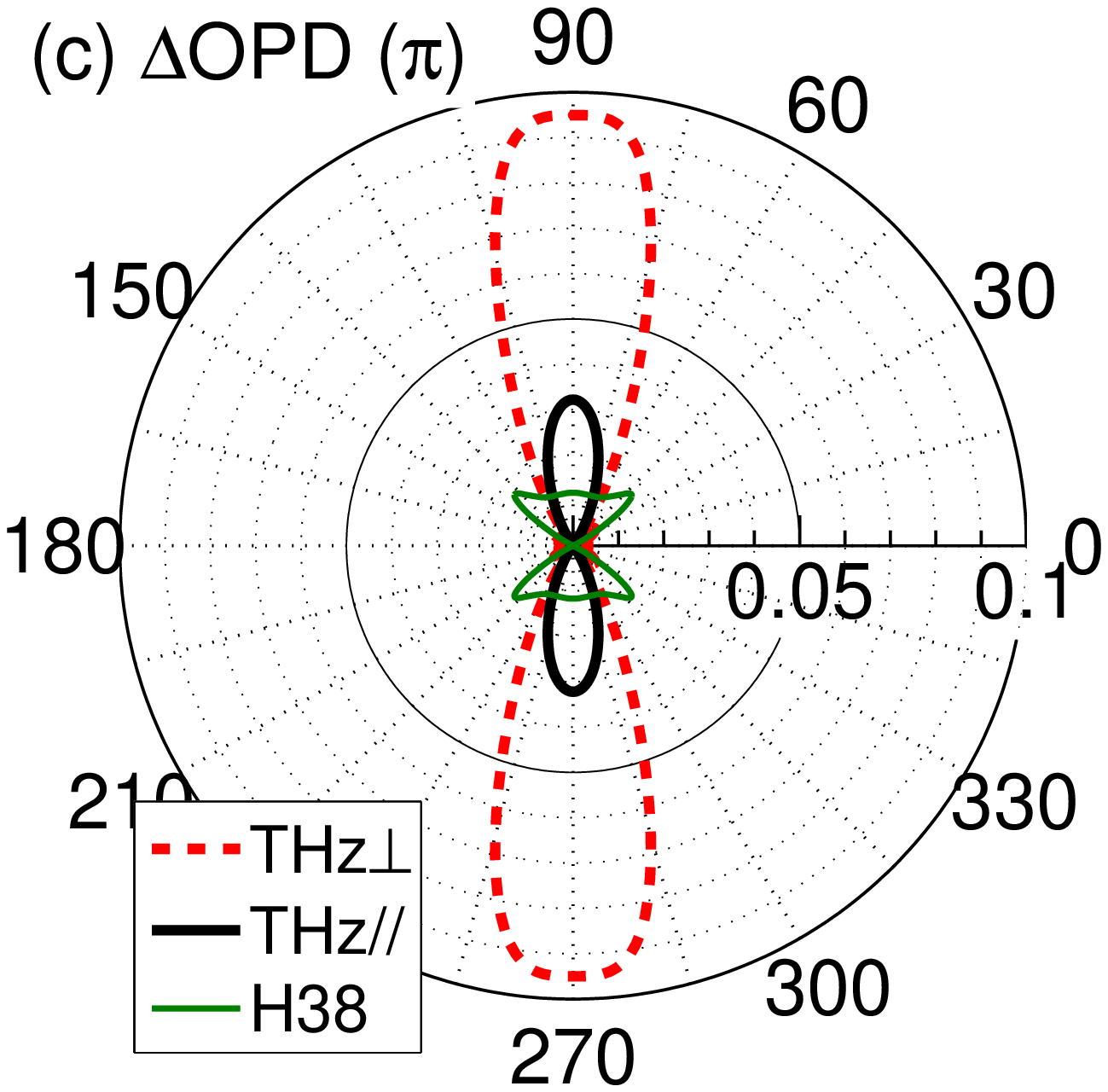}
\caption{\label{fig:molthz-i4-opd}(Color online) Alignment-resolved OPD and OTY for THz generation from H$_{2}^{+}$. (a) Normalized OTY parallel (black solid) and perpendencular (red dash) to laser polarization, ionization population (blue dash-dot) and the 30th HHG (thin blue solid). (b) OPD. (c) OPD difference at various alignment angles  to that of $\theta=0^\circ$. The laser pulse parameters are $\omega=0.057$au, $I_\omega=4I_0$, $I_{2\omega}=0.5\%I_\omega$, FWHM=15fs.}
\end{figure}

For given two-color laser intensity, e.g. $I_{\omega}=4\times10^{14}\mbox{W/cm}^{2}, I_{2\omega}=0.5\%I_{\omega}$, we examine in details how the OTYs vary with alignment angle. Figure \ref{fig:molthz-i4-opd}(a) shows that the total OTYs take maximum at $\theta=0^\circ$ and minimum at $\theta=90^\circ$ respectivley, which closely assembles the angular dependence of the ionization probability. However, the perpendicular component of THz wave, although much weaker, exhibits very different alignment dependence, reaching maximum at $\theta=45^\circ$ in contrast to the parallel component.

We now investigate the alignment dependence of the OPDs. As shown in Fig. \ref{fig:molthz-i4-opd}(b), the OPDs of the total THz wave yields (mainly the parallel component) varies with the alignment angle $\theta$. The maximum variation $\Delta\phi_{m}=\phi_{m}\left(\theta\right)-\phi_{m}\left(0\right)$ is found less than $0.1\pi$. Although the variation is small, but it indeed indicates that the molecular potential plays a role in the generation of THz waves. In fact, converting into time scale, the phase change of $0.1\pi$ corresponds to a time delay of 67 attoseconds (as). For better illustration, the variations $\Delta\phi_{m}$ are plotted for both the parallel and perpendicular components in Fig. \ref{fig:molthz-i4-opd}(c). For both components, it can be seen that the OPD increases as the the alignment angle increases and reaches maximum at $90^\circ$ alignment. The extra phase delay for molecules perpendicular to the laser polarization might be attributed to the larger recollision cross section in this geometry which leads to larger distortion of the electron trajectories and hence more pronounced focusing effects during THz wave emission.

\begin{figure}[t]
\includegraphics[clip,width=0.3\columnwidth]{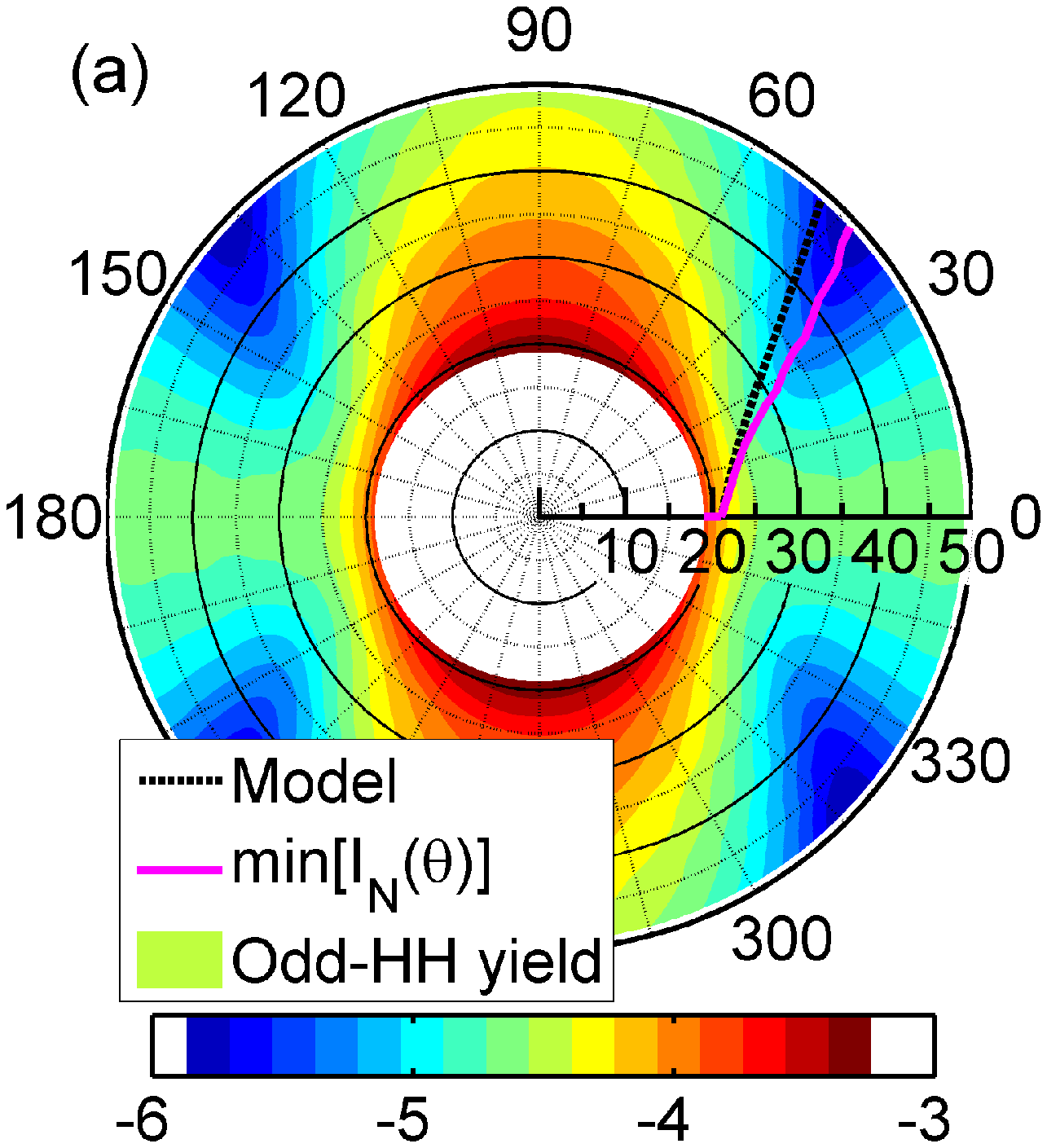}\,
\includegraphics[clip,width=0.3\columnwidth]{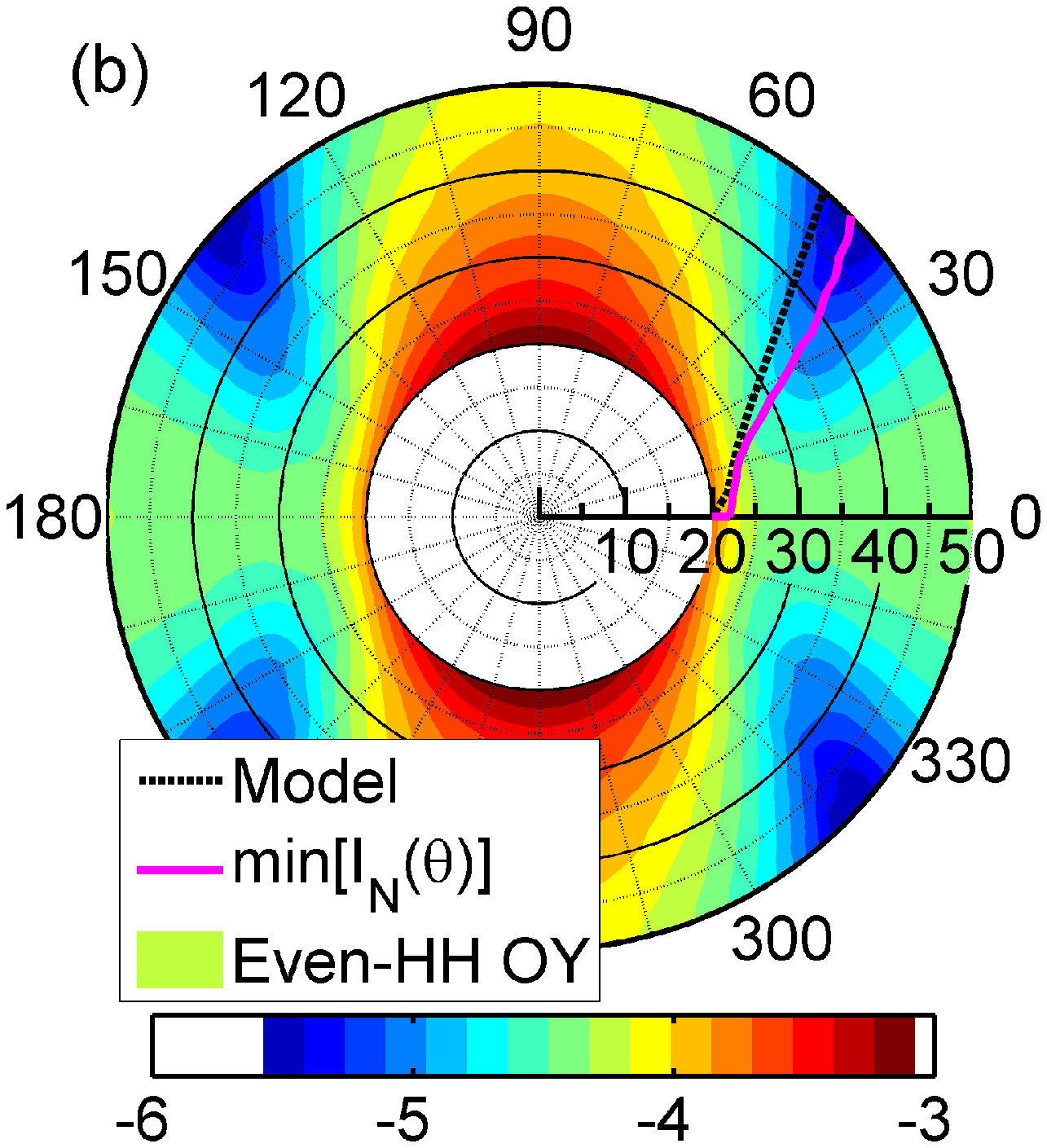}\,
\includegraphics[clip,width=0.3\columnwidth]{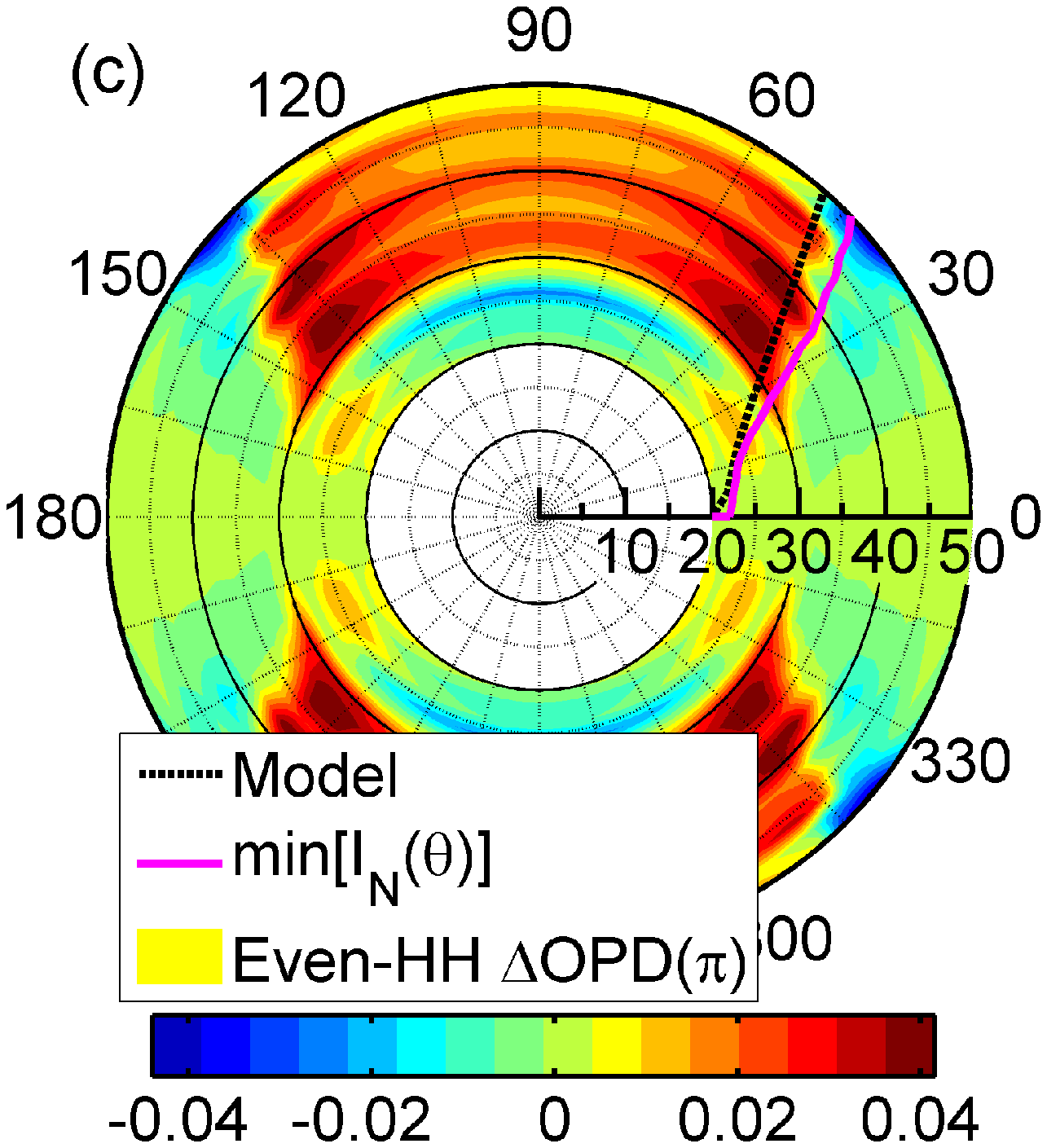}\\
\includegraphics[clip,width=0.32\columnwidth]{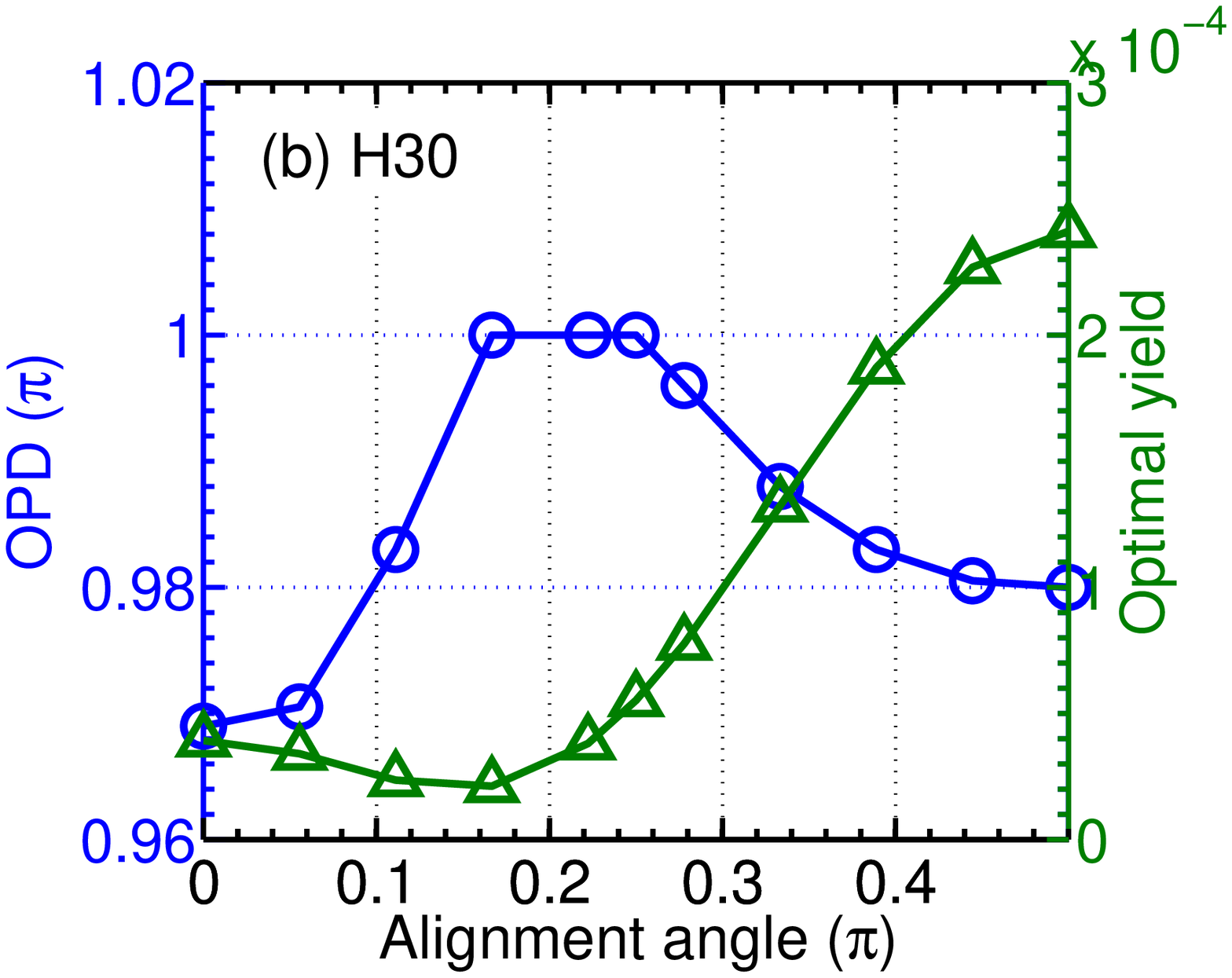}
\includegraphics[clip,width=0.32\columnwidth]{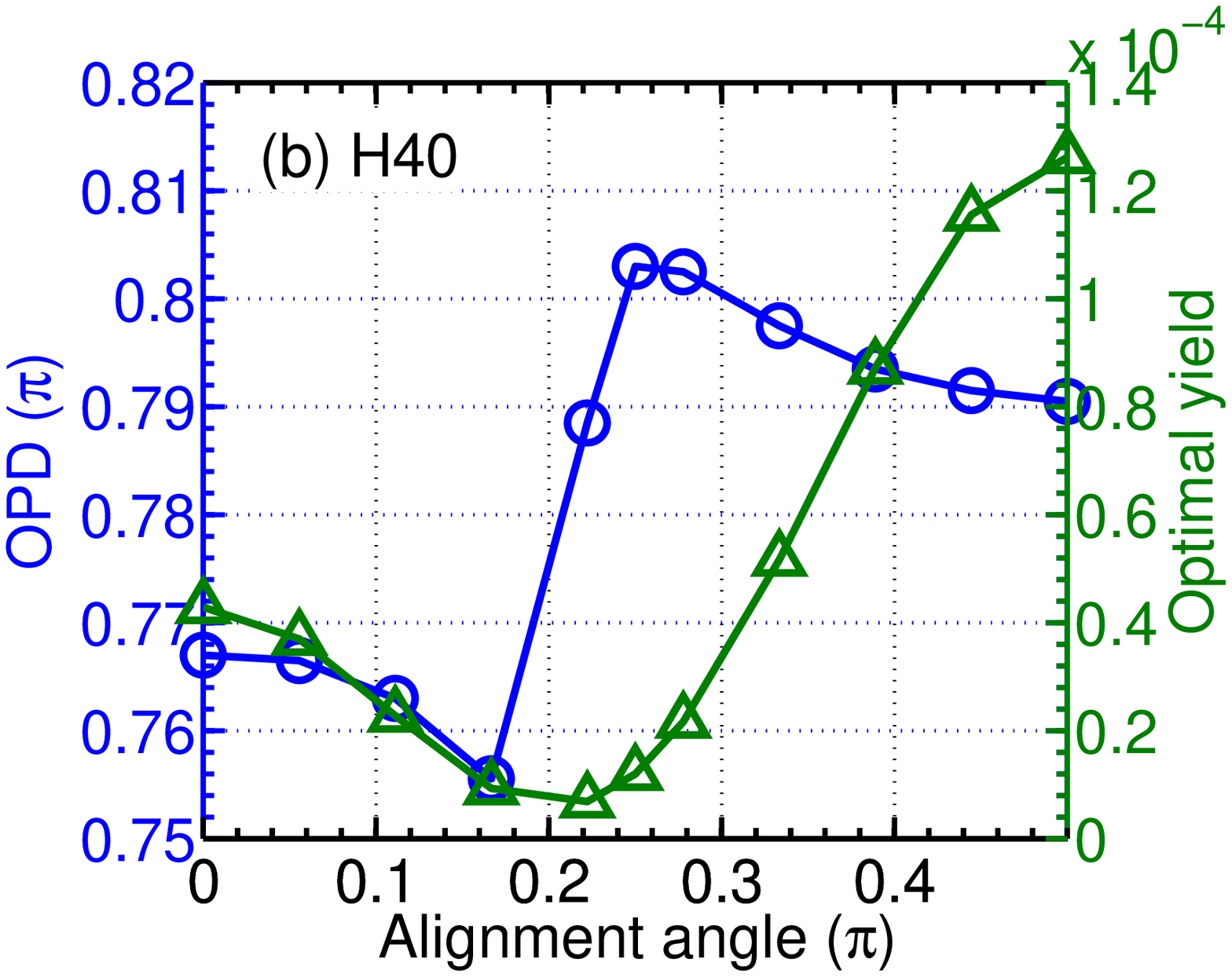}
\includegraphics[clip,width=0.32\columnwidth]{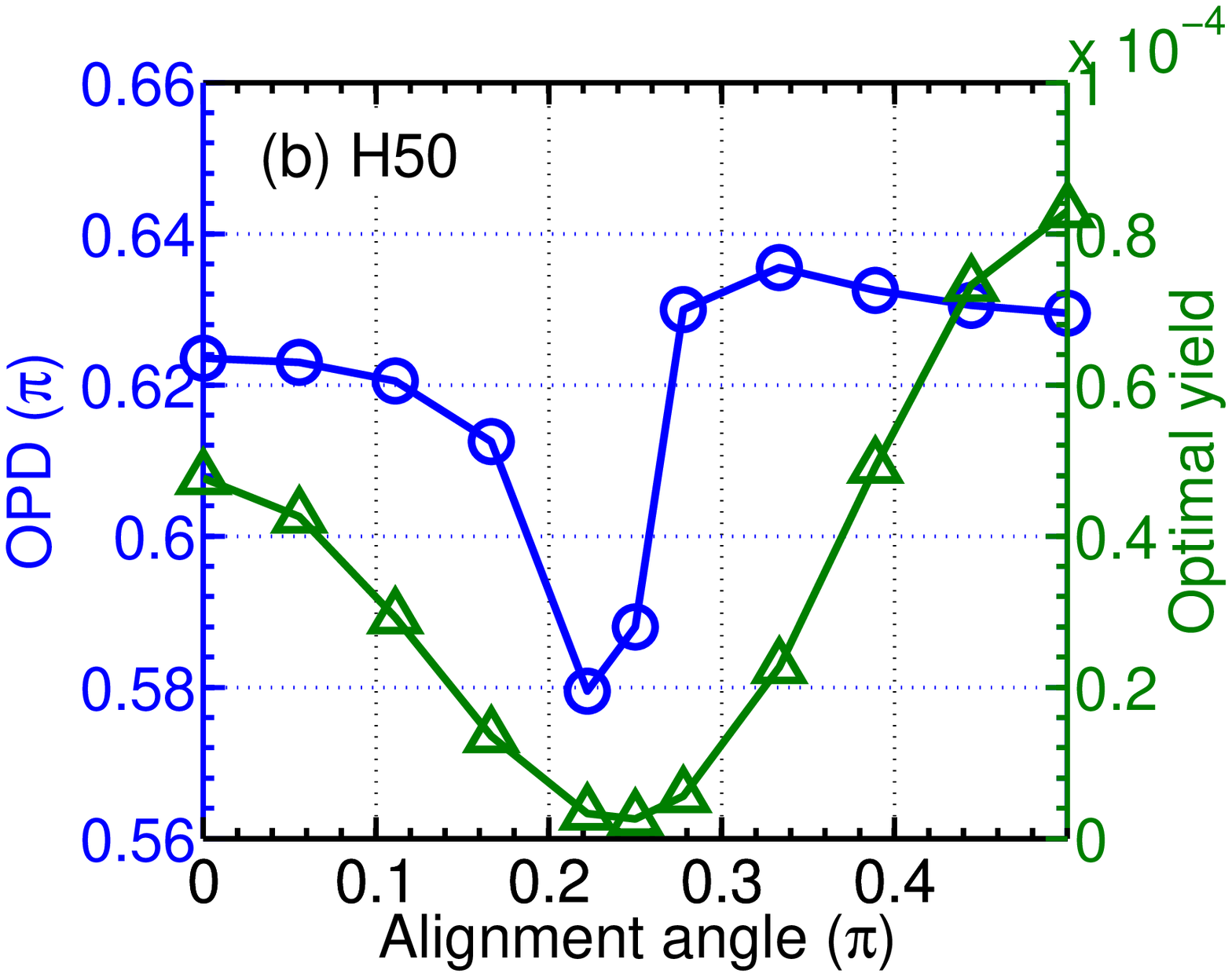}
\caption{\label{fig:molhhg-allangle}(Color online) Harmonic generation from H$_{2}^{+}$. (a) Odd harmonic yields in one-color laser. (b) Even harmonic  yields from the two-color laser pulse. (c) OPD difference of even  harmonics relative to $\theta=0^\circ$. (d), (e), (f), the OPDs and optimum yields of harmonics 30th, 40th and 50th as functions of on alignment angles $\theta$, respectively. }
\end{figure}

In order to provide more insight into the generation of terahertz waves from molecules, we present the results of high harmonic generation (HHG) as well. In Fig.~\ref{fig:molhhg-allangle}(b), the optimal yields of even harmonics from the two-color laser pulse are shown in contour plot against the alignment angles. For comparison, high harmonic yields from the fundamental pulse alone are presented in Fig.~\ref{fig:molhhg-allangle}(a). It can be seen that the yields of even harmonics generated by the two color laser field peak at both alignment of $0^\circ$ and $90^\circ$, which assembles the behavior of odd harmonics generated by the one-color laser pulse. However this is in very contrast to the alignment dependence of terahertz radiation yields which peak at zero alignment (see Fig. \ref{fig:molthz-i4-opd}(a) ). It is worth to note that the minimum of harmonic yields (shown in red lines in Fig.~\ref{fig:molhhg-allangle}(a) and (b)) are in accordance to the predication (shown in black lines) that the destructive two-center interference occurs for
\begin{equation}
R\cos\theta=\left(2m+1\right)\lambda/2,m=0,1,\ldots,
\end{equation}
where $\lambda$ is the electron wavelength, and electron energy is determined by $E_{k}=N\omega$ \cite{Lein02A2}.

By varying the phase-delay between the two pulses, both even harmonic and terahertz wave yields modulate. In Fig. \ref{fig:molhhg-allangle}(d-f), we show the OPD and OTY for 30th, 40th and 50the harmonics as functions of alignment respectively. It can be seen wherever the OTY takes the minimum, the corresponding OPD makes a phase jump. In contrast, the OPD of THz wave show a slow variation with the molecular alignment (see Fig. \ref{fig:molthz-i4-opd}(b)).

\begin{figure}[htdp]
\includegraphics[clip,width=0.32\columnwidth]{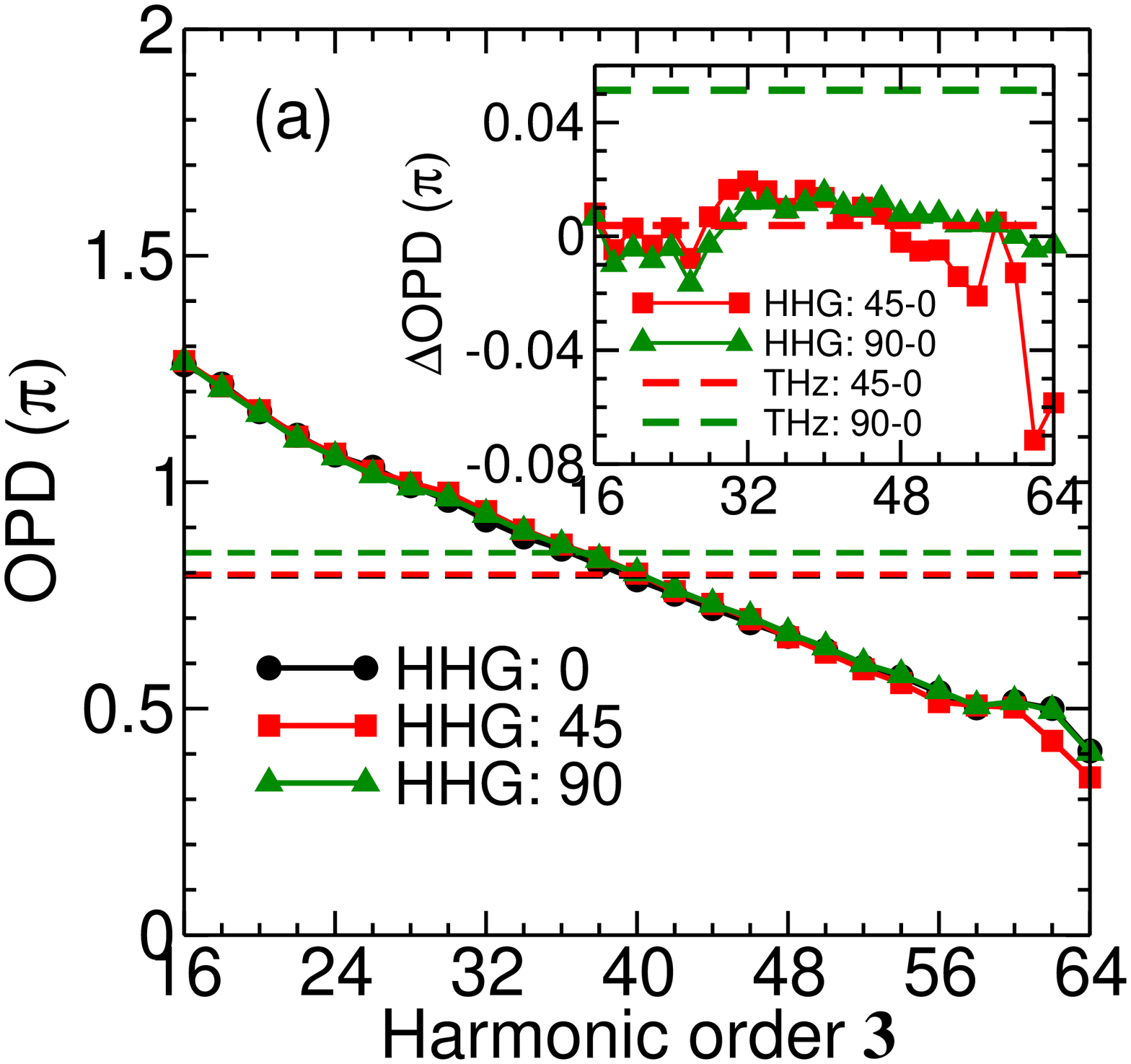}
\includegraphics[clip,width=0.32\columnwidth]{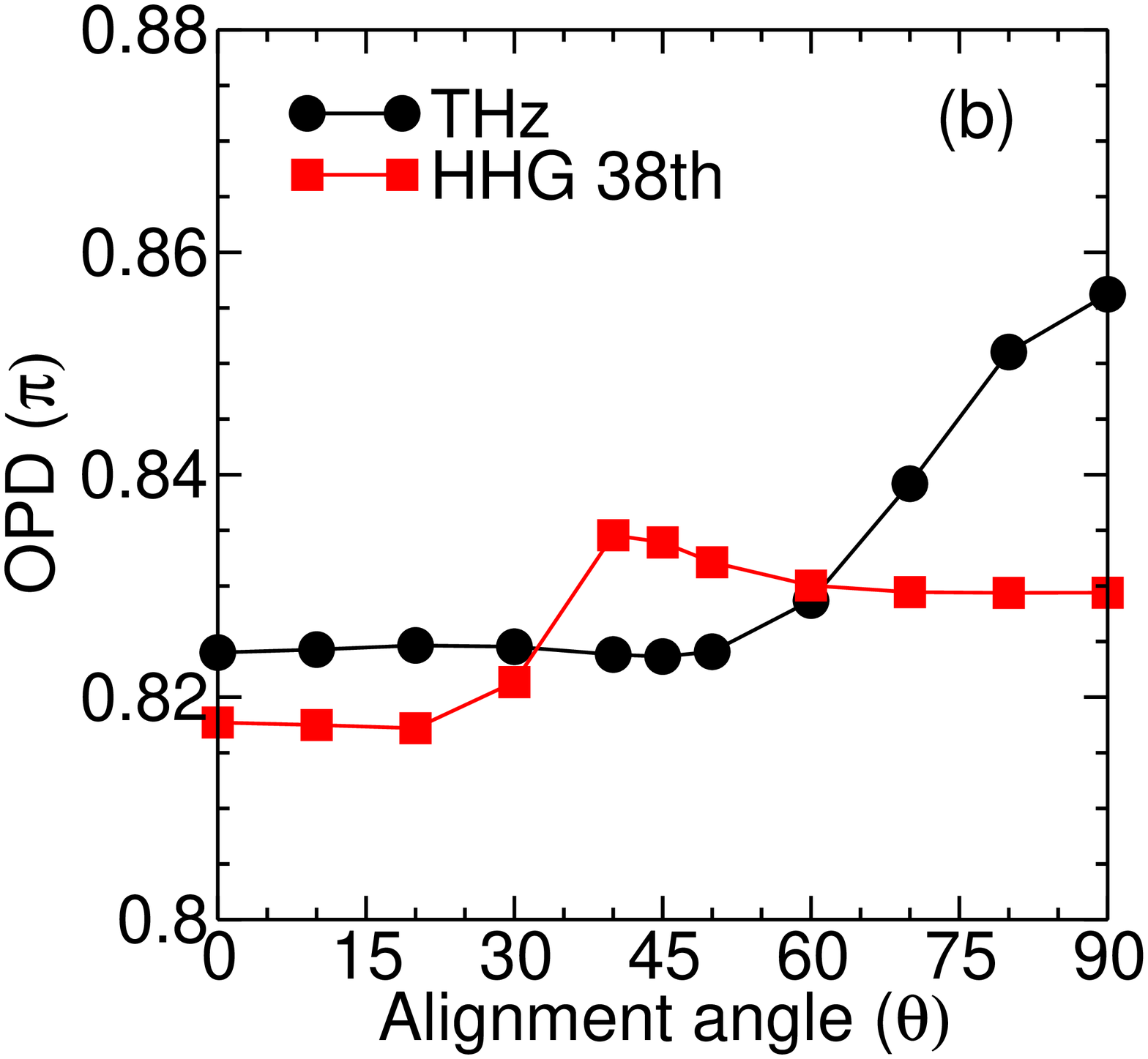}
\includegraphics[clip,height=0.3\columnwidth]{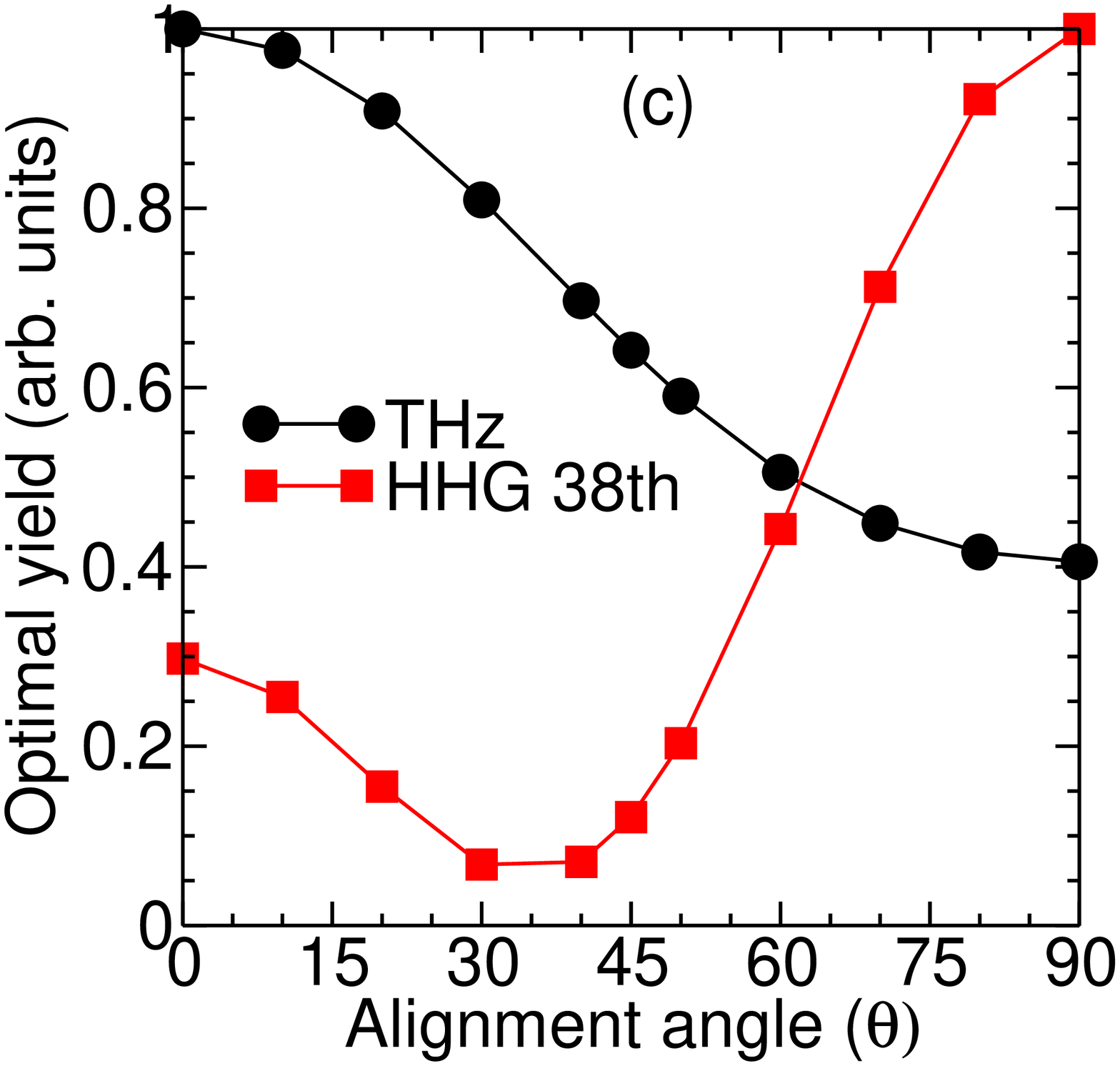}
\caption{\label{fig:molhhg-90d}(Color online) (a) The OPD of HHG at alignment angles of $0^\circ$, $45^\circ$ and $90^\circ$. The black, red, and green dashed lines denote the OPD of terahertz waves at the corresponding angles. The inset shows the difference of OPD at $45^\circ$ and $90^\circ$ with respect to the alignment of $0^\circ$ angle. The alignment-dependent  (b) OPDs and (c) optimum yields  of THz generation and  the yields of the 38th harmonics in the two-color laser pulse. }
\end{figure}

For alignment of $0^\circ$, $45^\circ$ and $90^\circ$, the OPD's as functions of the harmonic order are shown in Fig. \ref{fig:molhhg-90d}(a). It can be seen in all three cases that the OPD is nearly linear with respect to the harmonic order, reflecting the chirp of the recombine electron wave packet during HHG processes \cite{Mairesse03}. Detailed examination shows that the OPD for alignment $45^\circ$ and $90^\circ$ differs slightly from that of $0^\circ$ alignment.

To make more close contact of THz and HHG, the OPD and OTY of 38th harmonic yields are shown in comparison with terahertz generation in Fig. \ref{fig:molhhg-90d}(b) and (c). It clearly demonstrates that THz yields can be used to gauge the harmonic yield to gain the two-center interference information of HHG. The phase-delay dependence of THz generation can be used to probe the modulation of harmonics as well.

\section{Conclusions}

In conclusion, we have investigated the atomic and molecular terahertz generation with intense two-coloar laser pulses. Based on the TDSE simulation, we show that THz generations is dominated by the competition between laser field and long-range coulomb field, which can affect the ionization mechanism and wave-packets dynamics. As the laser intensity increases, the OPD varations demonstrate the dominated mechanisms of THz generation are changing from four-wave-mixing, rescattering currents of soft-recollision between ionized electron with atomic core, to photocurrent model without Coulomb potential. When we focus on THz generation dependence on both molecular alignment angle and molecular potential, the total OTY is closely similar to the angular dependence of the ionization probability, and the OPD variation of several tens of attoseconds indicates that the molecular potential plays a role in the generation of THz waves. Finally we show that both the alignment-dependence of the OTY and OPD of THz generation, can be used to contrast that of high harmonic generation, which might be used to explore further the corresponding attosecond electron dynamics experimentally.


\begin{acknowledgments}
This work is supported by the National Basic Research Program of China (973 Program) under Grant No. 2013CB922203, and the NSF of China (Grants No. 11374366).
\end{acknowledgments}


\end{document}